\documentclass[%
 reprint,
superscriptaddress,
 amsmath,amssymb,
 aps,
]{revtex4-2}

\usepackage{graphicx}
\usepackage{dcolumn}
\usepackage{bm}

\usepackage{physics}
\usepackage{upgreek}
\usepackage{gensymb}
\usepackage{units}
\usepackage{float}
\usepackage[colorlinks=true,linkcolor=blue,urlcolor=blue,citecolor=blue]{hyperref}

\usepackage{float}
\usepackage[commentmarkup=uwave,deletedmarkup=sout]{changes}
\definechangesauthor[name=Malte Grosche, color=blue]{fmg}
\begin{document}
\title{Metamagnetism in UTe$_2$: the roles of itinerancy and localization}

\author{Theodore I. Weinberger}
\affiliation{Cavendish Laboratory, University of Cambridge,\\
 JJ Thomson Avenue, Cambridge, CB3 0US, United Kingdom}

\author{Daniel Shaffer}
\affiliation{Department of Physics, University of Wisconsin-Madison, Madison, Wisconsin 53706, USA}

\author{Zheyu Wu}
\affiliation{Cavendish Laboratory, University of Cambridge,\\
 JJ Thomson Avenue, Cambridge, CB3 0US, United Kingdom}

\author{Dmitry V. Chichinadze}
\affiliation{National High Magnetic Field Laboratory, Tallahassee, Florida, 32310, USA}
\affiliation{Department of Physics, Washington University in St. Louis, St. Louis, MO, 63130, USA}

\author{Jinxu~Pu}
\affiliation{School of Physics and Astronomy, Shanghai Jiao Tong University, Shanghai, 200240, China}
\affiliation{Department of Physics, Clarendon Laboratory, University of Oxford, Parks Road, Oxford OX1 3PU, UK}

\author{Gang Li}
\author{Rui Zhou}
\affiliation{Beijing National Laboratory for Condensed Matter Physics, Institute of Physics, Chinese Academy of Sciences, Beijing 100190, China}
\affiliation{School of Physical Sciences, University of Chinese Academy of Sciences, Beijing 100190, China}

\author{Yurii Skourski}
\affiliation{Hochfeld-Magnetlabor Dresden (HLD-EMFL),\\
Helmholtz-Zentrum Dresden-Rossendorf, Dresden, 01328, Germany}

\author{Dave~Graf}
\affiliation{National High Magnetic Field Laboratory, Tallahassee, Florida, 32310, USA}

 \author{Andrej Cabala}
 \author{Vladim\'{i}r Sechovsk\'{y}}
 \author{Michal Vali{\v{s}}ka}
 \affiliation{Charles University, Faculty of Mathematics and Physics,\\ Department of
Condensed Matter Physics, Ke Karlovu 5, Prague 2, 121 16, Czech Republic}

\author{Michal P. Kwasigroch}
\affiliation{Department of Mathematics, University College London, Gordon St., London WC1H 0AY, United Kingdom}
\affiliation{Trinity College, Cambridge, CB2 1TQ, United Kingdom}

\author{F. Malte Grosche}
\author{Alexander G. Eaton}
 \email{alex.eaton@phy.cam.ac.uk}
\affiliation{Cavendish Laboratory, University of Cambridge,\\
 JJ Thomson Avenue, Cambridge, CB3 0US, United Kingdom}
 
\date{\today}

\begin{abstract}
\noindent
The metamagnetic transition in UTe$_2$ plays a key role in stabilizing two enigmatic field-induced superconducting phases. One of these phases (SC2) is truncated by the transition, lying directly below it, while the other (SC3) sits predominantly above it and appears to be stabilized because of it. While numerous pulsed field studies have examined this transition, comparatively few steady field experiments have investigated it. Here we report a suite of measurements of metamgnetism in UTe$_2$, at ambient pressure by torque magnetometry and extraction magnetometry techniques, and of the magnetoconductance under pressure. Our steady field measurements resolve a complex sub-structure within the transition, with separate features that possess different temperature evolutions, pointing to distinct contributions from itinerant and localized moments. The itinerant contribution might relate to a possible spin-density wave state. We theoretically model the evolution of Kondo and RKKY interactions and propose that the SC2 state is stabilized under pressure due to the collapse of magnetic anisotropy, leading to an enhancement of longitudinal spin fluctuations along the hard $b$ axis, which are pair-forming in the $p$-wave channel.

\end{abstract}

\maketitle
\begin{figure}[t!]
    \includegraphics[width=0.95\linewidth]{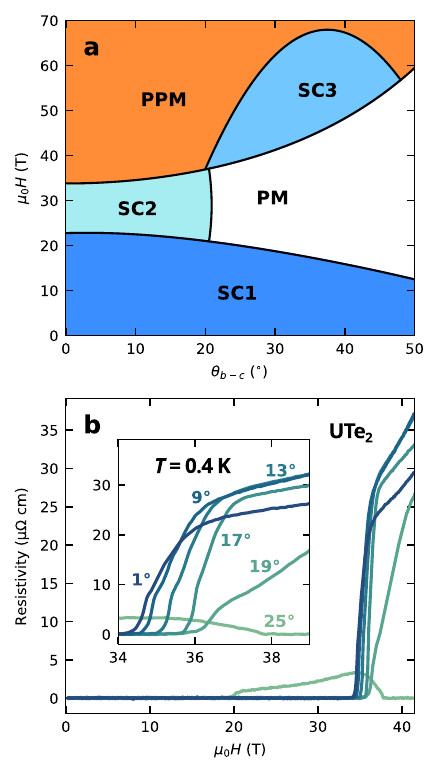}
    \caption{(a) Ambient pressure low temperature phase diagram of UTe$_2$ for magnetic field $H$ tilted at an angle $\theta$ in the crystallographic $b-c$ plane. For $H$ aligned close to $b$ ($\theta = 0\degree$), the ground state SC1 superconductivity transitions into a separate, distinct superconductive phase (SC2). The SC2 state is abruptly truncated at a first-order metamagnetic phase boundary, whereupon the material transitions into the polarized paramagnetic (PPM) state. Another superconducting phase (SC3) resides predominantly within the PPM at high $H$ and inclined $\theta$. Figure constructed from measurements on high-quality UTe$_2$ crystals reported in~\cite{tony2024enhanced,tony2025brief,pnas24data,pnas25data}. (b) Resistivity data measured for incremental $\theta$ at 0.4 K reproduced from \cite{tony2024enhanced,pnas24data}. For $\theta < 20^{\circ}$, zero resistivity persists up to the MM transition. In each of these curves, the MM transition exhibits a small anomalous kink feature (below 10~$\upmu \Omega$cm), indicative of two steps in the transition.}
    \label{fig:bc-phasediag}
\end{figure}

 
\noindent
The heavy fermion metamagnet UTe$_2$ possesses a plethora of emergent electronically ordered phases. Under various conditions of magnetic field orientation and applied pressure, up to six distinct superconductive states have been reported in this material, several of which exhibit characteristics of odd-parity pairing~\cite{Ran2019Science,Aoki2019,Aoki_UTe2review2022,lewin2023review,Ranfieldboostednatphys2019,Knebel2019,Braithwaite2019,Thomas2020,aoki2020multiple,LinNevidomskyyPaglione20,Aoki2021,ran2021expansion,Rosuel23,honda2023pressure,LANL_bulk_UTe2,tony2024enhanced,helm2024,tony25pressure}. The ground state of the system, at ambient pressure $p$ and zero applied magnetic field \textbf{H}, is proximate to two magnetic instabilities. One is located at a critical pressure of $p_c \approx$~15~kbar, beyond which incommensurate antiferromagnetic ordering is observed~\cite{KnafoPRX25,Thomas2020}. The other is a metamagnetic (MM) transition to a ferromagnetic-like field-polarized state~\cite{Miyake2019}. For $p$~=~0 this transition is found at $\mu_0 H_m \approx$~34~T for \textbf{H} aligned along the hard magnetic $b$-axis~\cite{lewin2023review} (see Fig.~\ref{fig:bc-phasediag}). Notably, as $p$ increases towards $p_c$, $H_m$ decreases~\cite{LinNevidomskyyPaglione20}. Attaining a deeper understanding of the interplay between these various magnetic interactions in UTe$_2$ likely holds the key to unravelling the exotic superconductive pairing mechanisms at play in this material.

At low temperatures for $p = H =0$, UTe$_2$ is paramagnetic, exhibiting no long-range magnetic order~\cite{Aoki_UTe2review2022}. Given the numerous similitudes between UTe$_2$ and the  uranium-based ferromagnetic superconductors UGe$_2$, URhGe and UCoGe~\cite{Montu_UGe2,URhGE_Aoki2001,UCoGe_PhysRevLett.99.067006,Aoki_ferro_review2019}, initial studies of superconductivity in UTe$_2$ posited that ferromagnetic fluctuations may provide the pairing mechanism for spin-triplet superconductivity~\cite{Ran2019Science,Aoki2019,LinNevidomskyyPaglione20,RanPRB2020}. However, subsequent neutron scattering measurements reported only antiferromagnetic fluctuations with incommensurate wavevectors~\cite{DuanPRL2020,Duan2021AFMfluc,Knafo104.L100409}. Although early muon spin relaxation experiments resolved ferromagnetic fluctuations coexisting with superconductivity~\cite{UTe2-muons-PhysRevB.100.140502}, subsequent comparable measurements on higher quality samples did not reproduce these results~\cite{AzariPRL.131.226504}. This has led to proposals for triplet pairing mediated by spin fluctuations with high momentum ordering vector $Q$~\cite{Duan2021AFMfluc,KreiselPhysRevB.105.104507}, as opposed to the $Q=0$ scenario of ferromagnetic fluctuations.

By contrast, as $H$ is increased towards $H_m$, signatures of strong $Q = 0$ fluctuations emanating from the MM transition have been inferred from measurements of nuclear magnetic resonance (NMR)~\cite{Kinjo23,tokunaga2023longitudinal} and of the magnetic entropy~\cite{TokiwaPRB2024}. Bulk-sensitive thermodynamic measurements for \textbf{H}~$\parallel b$ have indicated the existence of a second-order phase boundary between two distinct superconducting phases for $\mu_0 H \approx$~15~T at low temperatures~\cite{Rosuel23}. As $\mu_0 H$ is increased beyond 15~T, NMR measurements find that longitudinal $Q = 0$ spin fluctuations grow in strength, before diverging as $H_m$ is approached~\cite{tokunaga2023longitudinal}. A rotation of the preferred spin orientation \textbf{S} of the putative spin-triplet Cooper pairs has also been proposed, from \textbf{S}~$\parallel a$ in the low field (SC1) state to \textbf{S}~$\parallel b$ in the higher field (SC2) phase~\cite{Kinjo23,tokunaga2023longitudinal}. For 4~kbar~$< p < p_c$, SC2 is stabilized at $\mu_0 H = 0$~T~\cite{VasinaPRL25}, with NMR signatures consistent with the scenario of the lower temperature SC1 state potentially having \textbf{S}~$\parallel a$ with the higher temperature SC2 phase possessing \textbf{S}~$\parallel b$~\cite{Kinjo_SciAdv23}. An outstanding challenge lies in building a unified understanding of how $H$ and $p$ each act to tune UTe$_2$ between the distinct SC1 and SC2 superconducting phases.

In our prior work~\cite{tony2024enhanced} we explored how crystalline disorder strongly suppresses the SC2 state. We constructed a simple model considering disorder-induced damping of $Q = 0$ MM fluctuations, which we found to accurately capture the experimental observations. Here we extend our study of the interplay between SC1, SC2 and metamagnetism as a function of hydrostatic pressure. First, we present high-$H$ measurements of UTe$_2$'s superconducting and magnetic phases over the interval of 0~kbar~$\leq p \leq$~25~kbar. Then we expand our theoretical model from ref.~\cite{tony2024enhanced} to investigate the effect of pressure on the Kondo and RKKY couplings. We find that considering the interplay between local and itinerant moments yields valuable insight into the nature of the MM transition and, by extension, shines light onto the character of the enigmatic SC2 state.

\section*{Metamagnetism at ambient pressure}

\begin{figure}[t!]
    \includegraphics[width=0.93\linewidth]{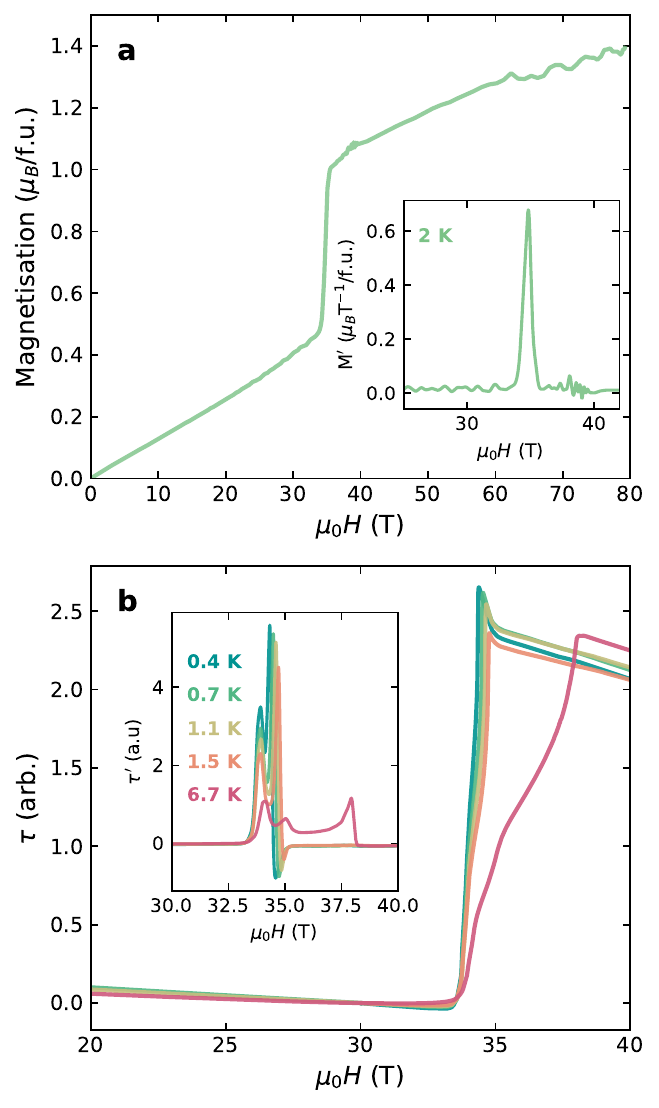}
    \caption{(a) Magnetization $M$ as a function of magnetic field $H$ measured by extraction magnetometry up to 80~T in a double-coil pulsed magnet system at HLD-EMFL. A sharp jump at $\mu_0H_m = 34$~T is observed, likely due to abrupt polarization of the localized uranium moments~\cite{Aoki_UTe2review2022}. The continued rise in $M$ post-transition indicates the presence of unsaturated itinerant moments. (Inset) A single sharp transition is resolved in the derivative of $M$ with respect to $H$. (b) Magnetic torque $\tau$ as a function of $H$ for \textbf{H}~$\parallel b$. The transition profile displays two structural features with distinct thermal evolutions, implying separate contributions from localized and itinerant moments. (Inset) $\partial \tau/\partial H$ at 0.4~K resolves two maxima within the transition, revealing microscopic structure that broadens into three maxima by 6.7~K.}
    \label{fig:magnetisation}
\end{figure}

The magnetic phase accessed at $\mu_0 H >$~34~T for $p=0$, \textbf{H}~$\parallel b$, is often referred to as a ``polarized paramagnet" (PPM) or ``field-polarized state"~\cite{Aoki_UTe2review2022,Ranfieldboostednatphys2019,lewin2023review}. This is due to a large discontinuous increase in the magnetization \textbf{M} of approximately 0.5 Bohr magnetons per unit cell upon crossing the first-order MM transition boundary~\cite{Miyake2019,Miyake2021}. Notably, however, after the MM transition $M$ still rises with increasing $H$~\cite{Miyake2019,Miyake2021}, as visible in Figure~\ref{fig:magnetisation}a. This suggests that, while there is good evidence for an abrupt saturation of the local uranium-site moment at $H_m$~\cite{Aoki_UTe2review2022}, as $\nicefrac{\partial M}{\partial H}$ is similar on either side of $H_m$ the itinerant contribution to $M$ is therefore still considerable within the PPM state -- even as high as $\mu_0 H =$~80~T. This could have important implications for the field-induced SC2 and SC3 phases.

Signatures of possible distinctions between the contributions from localized and itinerant moments might be resolved through torque magnetometry measurements~\cite{Griessen13.375}. This technique is very sensitive to small changes in the anisotropy of the magnetic susceptibility tensor. This is because the magnetic torque $\boldsymbol{\tau} = \mathbf{M} \cross \mathbf{B}$ where $\mathbf{B} = \mu_0(\mathbf{M} + \mathbf{H})$. $\boldsymbol{\tau}$ is therefore insensitive to the magnitude of $\mathbf{M} \parallel \mathbf{B}$ -- which in UTe$_2$ is very large for $B \geq \mu_0H_m$ -- but is highly sensitive to the deviation in orientation of \textbf{M} away from that of \textbf{B}. By way of example, magnetic torque measurements have previously been used to discern relative changes between orthogonal components of the magnetic susceptibility tensor down to one part in 10$^8$~\cite{sato2017thermodynamic,murayama2019diagonal}, highlighting the sensitivity of this technique to small changes in the anisotropy of \textbf{M}.

We measured the magnetic torque of a UTe$_2$ single crystal by the capacitive torque magnetometry technique with \textbf{H}~$\parallel b$, for which the data are plotted in Figure~\ref{fig:magnetisation}b. These measurements were performed on the same sample that in our prior quantum oscillation study~\cite{Eaton2024} exhibited quantum oscillatory frequencies up to 18.5~kT, indicative of very high crystalline quality. The MM transition is clearly visible as a sudden jump in $\tau(H)$. However, upon close inspection of the data, distinct features within the transition are visible. Over the temperature ($T$) range of 0.4~K to 1.5~K, the initial turning up of $\tau$ just below 34~T is largely insensitive to this temperature increment. However, between 34.0~T to 34.5~T there is a clear variation in $\tau(H,T)$ (see inset to Fig.~\ref{fig:magnetisation}b), with the location of the maximum of $\tau$ moving to higher $H$ with increasing $T$. At 6.7~K this becomes especially pronounced, with the MM transition having been stretched out over the range of 34~T~$\lessapprox \mu_0H \lessapprox$~38~T. This structure within the MM transition -- of an initial upturn in $\tau$ that has lower sensitivity to $T$ than the subsequent broadening out before $\tau(H)$ reaches its peak -- points to different microscopic contributions to \textbf{M}. These might stem from the separate responses from itinerant and localized moments upon crossing the MM phase boundary. We posit that the relatively $T$-insensitive initial upturn in $\tau$ likely corresponds to the polarization of the local uranium moments, in turn affecting the itinerant contribution to \textbf{M}, which may exhibit higher sensitivity to small changes in $T$ due to the heavy fermion character of this material. 

\section*{Pressure-tuned magnetic and superconducting phases}

\begin{figure*}[t!]
\includegraphics[width=0.9\linewidth]{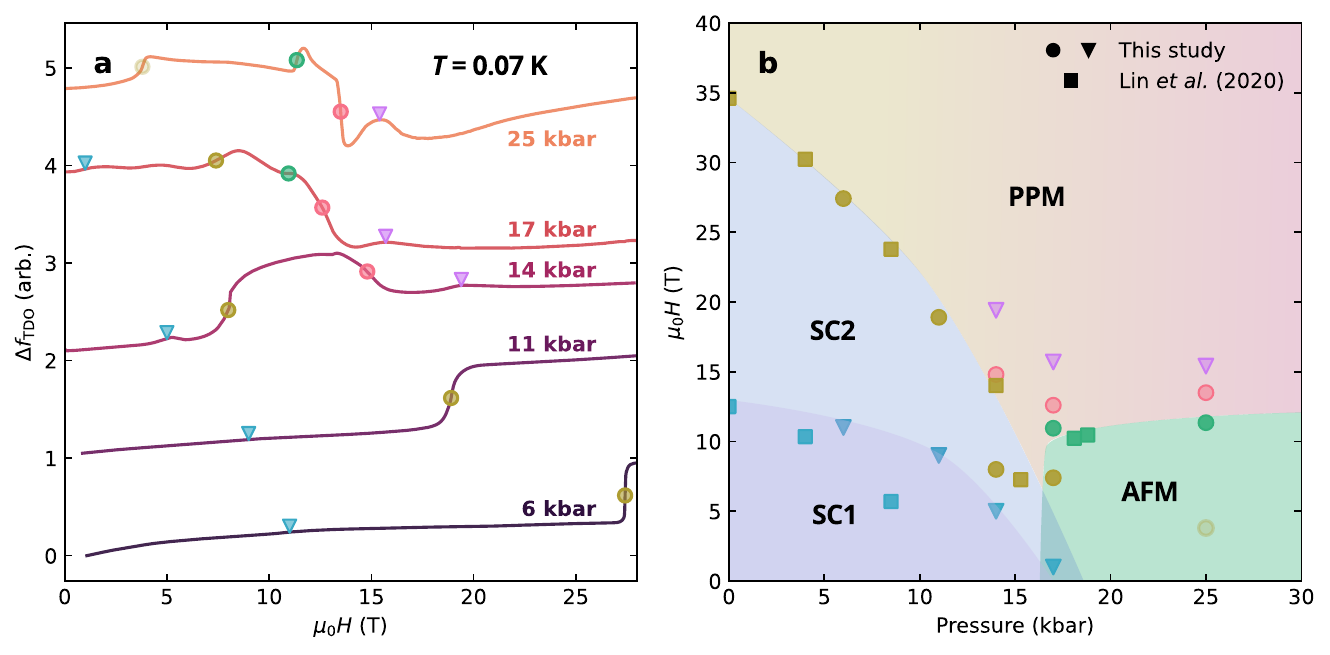}
    \caption{(a) Contactless resistivity measurements of UTe$_2$ under pressure for \textbf{H}~$\parallel b$. At 6~kbar we resolve both the transition between the SC1 and SC2 states (blue triangle) as well as a stark jump in the measured resonance TDO frequency at $H_m$ as the PPM state is accessed (brown circle). Increasing $p$ suppresses both of these transitions to lower $H$. For $p \geq$~14~kbar and $H > H_m$, a number of features can be seen in the contactless conductivity. We denote the transition demarcating the upper field of the high-pressure antiferromagnetic order by a green circle, beyond which the PPM state is restored. The nature of the other two features (marked with red circles and violet triangles) is unclear. At 25~kbar, we note that there is a sharp change in the TDO frequency at $\approx4$~T, which bears similarity to the jump seen at the onset of PPM, although previous work suggests the metamagnetic transition should occur at much higher fields \cite{lewin2023review}. (b) Low temperature $H-p$ phase diagram of the superconducting and magnetic states of UTe$_2$ for \textbf{H} $\parallel b$. Square points are taken from \cite{LinNevidomskyyPaglione20} with all other points from this study. Solid colored areas are illustrative of the approximate regions occupied by each phase, deduced from the data in panel (a). We note that above approximately 3~kbar, the low field superconducting state may be distinct from that found at ambient pressure~\cite{Kinjo_SciAdv23,tony25pressure,kamat2026thermodynamicdiscoverytetracriticalityemergent}.}
    \label{fig:pressures}
\end{figure*}

\begin{figure*}[t!]
    \includegraphics[width=1\linewidth]{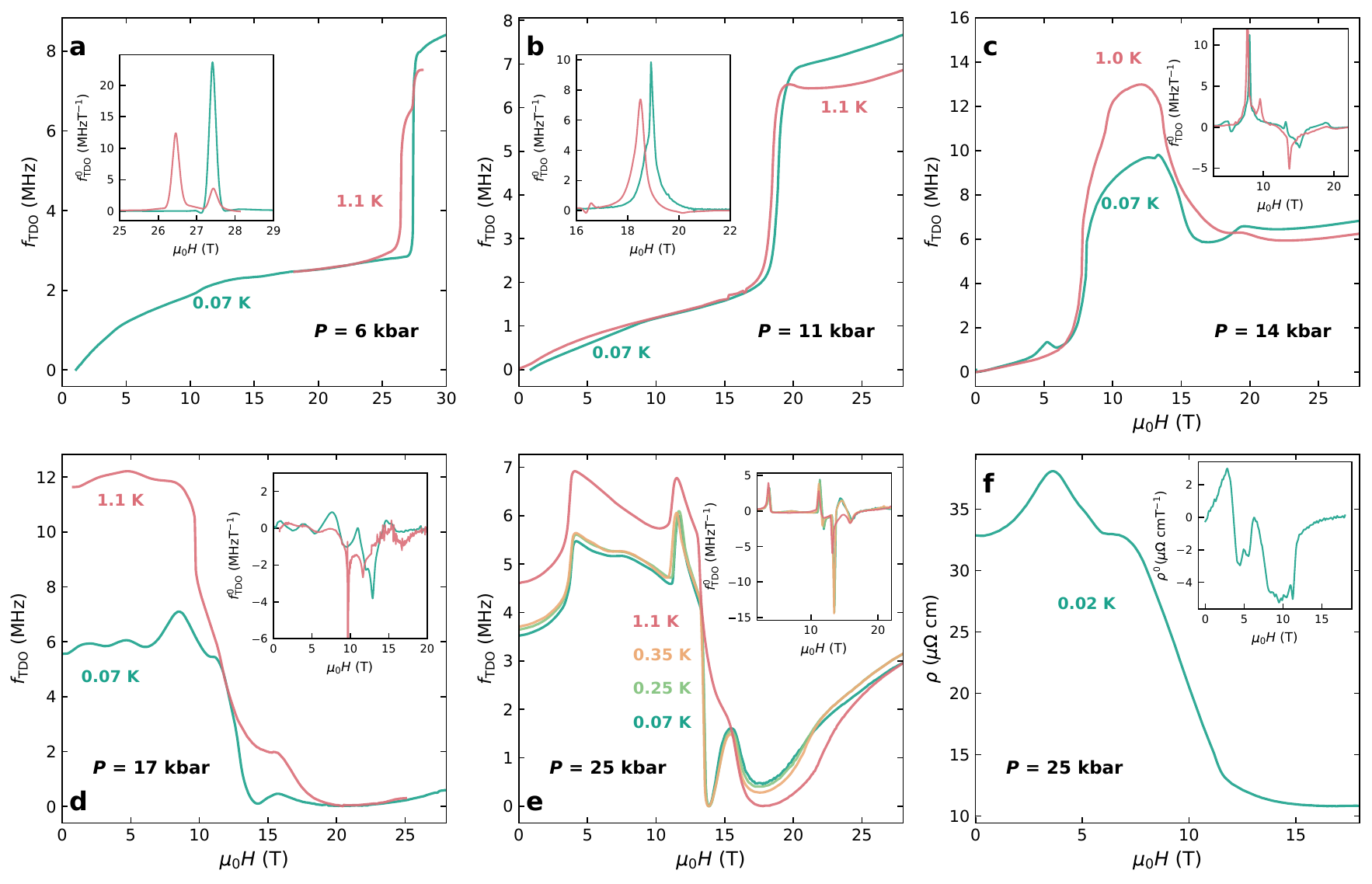}
    \caption{(a) Contactless resistivity of UTe$_2$ measured by the TDO technique for $p =$ 6~kbar at 70~mK and 1.1~K with \textbf{H}~$\parallel b$. At the lower temperature, the metamagnetic transition at $H_m$ manifests as a singular sharp step; whereas, upon warming it exhibits a two-step structure, clearly visible in the derivative plotted inset. Note that throughout this Figure curves have been translated to 0~MHz, for ease of presentation. (b) At 11~kbar $H_m$ has come down to 19~T, with the 1.1~K curve showing a small dip immediately above $H_m$. (c) At 14~kbar the dip following $H_m$ is pronounced even at low $T$, with these features now spread over several teslas. (d) At 17~kbar interesting features are exhibited within the AFM phase at 70~mK, which are much less pronounced at 1.1~K. The PPM state then appears to possess lower resistivity than is observed in the AFM phase. (e) At 25~kbar the profile of $f_\text{TDO}(H)$ is surprisingly complicated, pointing to the presence of multiple magnetic sub-phases not previously identified. Given that the AFM ordering is incommensurate~\cite{KnafoPRX25}, it is conceivable that it may undergo significant evolution(s) in field. (f) Here we plot contacted four-terminal electrical transport measured on a different sample at $P$ = 25~kbar, $T$ = 20~mK. Interestingly, the profile of this data is more similar to the low temperature TDO curve at 17~kbar than at 25~kbar. Given that we expect our pressure certainty to be good to within $\approx$ 1~kbar, this implies that the anomalous features observed in panel (e) either only occur after crossing a boundary somewhere within the range 24.5-25.5~kbar (that perhaps was not crossed in the $\rho$ measurement), or that these features relate to magnetic properties that TDO is sensitive to but yield a negligible impact on $\rho$.}
    \label{fig:4}
\end{figure*}

The $p-H-T$ phase landscape of UTe$_2$ has been extensively investigated through experiments on samples grown by the chemical vapour transport method, generally in magnetic fields up to $\approx$~35~T at pressures up to $\approx$~20~kbar~\cite{Aoki_UTe2review2022,lewin2023review,Braithwaite2019,Valiska2021,LinNevidomskyyPaglione20,ran2021expansion,aoki2020multiple,Thomas2020,tony25pressure,weinberger2025pressure}. Here we present measurements on a high quality molten salt flux-grown crystal \cite{PhysRevMaterials.6.073401MSF_UTe2} up to $\mu_0H$~=~30~T, $p$~=~25~kbar, recorded in a dilution refrigerator reaching $T <$~0.1~K.

In our prior comparative study of CVT and MSF UTe$_2$ specimens~\cite{tony2024enhanced}, at ambient pressure, we found that although the SC1 and SC2 phases are acutely sensitive to the presence of crystalline disorder, the MM transition is remarkably insensitive. This was evidenced by both generations of crystals showing an extremely similar profile of the MM transition in terms of both magnetic field strength and tilt angle. Here we present contactless resistivity measurements for \textbf{H}~$\parallel b$ at incremental pressures up to 25~kbar, plotted in Figure~\ref{fig:pressures}a. These data were acquired by the tunnel diode oscillator (TDO) \cite{RevSciIns_TDO} technique, following the procedure of \cite{theo2024,weinberger2025pressure}. In Figure~\ref{fig:pressures}b, we compare the evolution of the $H-p$ phase diagram as reported from prior measurements on a CVT sample \cite{LinNevidomskyyPaglione20} with our measurements here on an MSF specimen.


There are a number of interesting features in these UTe$_2$ magnetoconductivity data. In Figure~\ref{fig:4} we plot the 70~mK traces from Fig.~\ref{fig:pressures}a alongside higher temperature data at each pressure increment. At 6~kbar and 11~kbar, there are subtle differences in the form of the metamagnetic transition between 70~mK and 1.1~K, reminiscent of the change in profile of $\tau(H,T)$ at $H_m$ in Fig.~\ref{fig:magnetisation}b. For $p \geq$~14~kbar, the form of $f_\text{TDO}(H)$ becomes very complicated. At $p =$~14~kbar, $T =$~70~mK, there is a small feature at 5~T signifying the SC1-SC2 boundary. $f_\text{TDO}$ then rises sharply around 7~T, characteristic of the sample resistivity increasing. The signal then starts to plateau, before falling around 15~T, with another small feature at 19~T. Upon warming to 1.0~K, the general profile of $f_\text{TDO}(H)$ is retained, but the 19~T feature is smeared out and that at 5~T is absent.

At $p = 17$~kbar, $T = 70$~mK (Fig.~\ref{fig:4}d) there are a number of hump-like features within the AFM state. These have previously been observed in low temperature measurements of the $c$-axis resistivity in a CVT sample at a similar pressure~\cite{KnebelPRBcaxis24}. Their precise origin is unclear, but suggestive of complex spin reorientations within the AFM state~\cite{Valiska2021}. Upon warming to 1.1~K these features become much smoother, while the drop in $f_\text{TDO}(H)$ upon exiting the AFM state becomes much more pronounced.

The only previous study measuring the TDO response of UTe$_2$ under pressure for \textbf{H}~$\parallel b$ went to a maximum pressure of 19~kbar~\cite{LinNevidomskyyPaglione20}. One might expect the profile of $f_\text{TDO}(H)$ to become quite simple at yet higher pressures, as the material is pushed further away from $p_c$. However, at $p =$~25~kbar we resolve a number of anomalous features in the TDO signal (Fig.~\ref{fig:4}e). There are two steep upturns in $f_\text{TDO}$ at 3.8~T and 11.3~T, before a very sharp decrease at 13.5~T. This is followed by further non-monotonicity of $f_\text{TDO}(H)$ until 20~T where the slope is positive with increasing $H$ up to the highest measurement field of 29~T. Notably, such a profile is not observed in contacted transport (panel f). How these features might relate to recent observations of possible magnetic subphases within the PPM at ambient pressure~\cite{ripples} is presently unclear.

The temperature dependence of $f_\text{TDO}(H)$ yields insight into the relative energy scales of these features in the contactless conductivity. At 17~kbar, following the downturn around 14~T there is a small local maximum in $f_\text{TDO}(H)$ around 16~T at 70~mK, which upon warming to 1.1~K becomes much less pronounced. At $p =$~25~kbar the temperature evolution is even starker, with the local maximum around 15~T for $T \leq$~0.35~K entirely washed out by 1.1~K. This points to a very low energy scale of this anomalous feature in $f_\text{TDO}(H)$ at high $p$.

\section*{Evolution of Kondo interactions}

In our earlier work \cite{tony2024enhanced}, we proposed that the SC2 state of UTe\(_2\) is closely connected to the MM transition that occurs at a critical value \(H_m\) of the magnetic field applied along the \(b\) axis. Experimentally, it has been observed here and in the literature that when pressure is applied, \(H_m\) decreases while a phase transition occurs from SC1 to SC2, with \(T_c\) growing in the SC2 phase~\cite{LinNevidomskyyPaglione20,lewin2023review,tony25pressure,VasinaPRL25}. This is consistent with the model for SC2 introduced in \cite{tony2024enhanced}, which, in the absence of an applied field and disorder, is given by
\begin{equation}
    T_c^{\mathrm{(SC2)}}=1.13\Lambda \exp\left[-\frac{H_m^2}{8\tilde{\nu}  M_*^2}\right],
    \label{TcSC2}
\end{equation}
where \(M_*\) is the magnetization above the metamagnetic phase transition and $\tilde{\nu}$ is proportional to the density of states. In particular, \(T_c^{\mathrm{(SC2)}}\) grows as \(H_m\) decreases.

The decrease of \(H_m\) with applied pressure, in turn, can be accounted for using a Kondo lattice model with uranium \(f\) electrons in the role of localized moments \(\mathbf{S}_{i,f}\) (\(i\) being a lattice site index) coupled to delocalized uranium \(d\) and tellurium \(p\) electrons via the Hamiltonian
\begin{equation}
    H_K= \sum_i J_K \vec{S}_{i,f} \cdot \vec{s},
\end{equation}
where \(J_K>0\) is the Kondo coupling and \(\mathbf{s}=\sum_{\mathbf{p}s_1s_2}c^\dagger_{\mathbf{p}s_1} \boldsymbol{\sigma}_{s_1s_2}c_{\mathbf{p}s2}\) is the spin of the mobile electrons. This model exhibits a metamagnetic phase transition when a Zeeman term \(H_Z=2 \mu_e \mathbf{H}\cdot \mathbf{s}\) is included, as has been seen for example in \cite{Burdin09, Bernhard22, Thomas23}. It is found in general that the applied field needs to exceed the value set by the Kondo scale to induce a first-order phase transition from the Kondo phase into the spin-polarized phase. In particular, \(H_m\) increases with increasing \(J_K\) and behaves in the same way as \(J_K\) under pressure.

Moreover, the behavior of \(J_K\) under pressure is in general a complicated question: it is known to either increase or decrease with pressure in different materials, for example, in cerium-based and ytterbium-based Kondo materials, respectively \cite{Takabatake98, Goltsev05}. 
In simple terms, this can be understood from the textbook expression for the Kondo coupling
\cite{Altland2010book,Coleman2015book}:
\begin{equation}
J_K = V^2 \left[ \frac{1}{U + \epsilon_f} - \frac{1}{\epsilon_f} \right],
\end{equation}
where $V$ describes hybridization (coupling) between the electrons on the local impurity site and the itinerant (mobile) electrons, $\epsilon_f$ is the negative chemical potential of electrons on the impurity site, and $U$ is the Hubbard-like electron-electron interaction between itinerant electrons. Of these, \(U\) is not expected to depend on pressure as it describes on-site interactions sensitively, but both \(V\) and \(\epsilon_f\) in general 
are more sensitive to
pressure. Generally \(V\propto e^{-r\mathcal{E}}\) where \(r>0\) and \(\mathcal{E}\) is strain, and so increases with pressure \cite{Goltsev05}. \(\epsilon_f\), on the other hand, may increase or decrease with pressure. The latter is seen in ytterbium-based Kondo materials and has been attributed to negative valence fluctuation in ytterbium (in contrast to cerium, which has positive valence fluctuations).

It has been additionally shown that the Gr\"uneisen parameter, \(\Omega=-\frac{d \ln T_K}{d \ln \mathcal{V}}\) where \(T_K\) is the Kondo temperature and \(\mathcal{V}\) is the volume, can become negative when \(J_K\) decreases with pressure \cite{Goltsev05}. Crucially, the Gr\"uneisen parameter measured for the putative Kondo temperature \(T_*\) is negative in UTe\(_2\) \cite{Aoki_UTe2review2022}. This suggests that valence fluctuations likely play an important role in UTe\(_2\), as has been extensively discussed in the literature~\cite{Aoki_UTe2review2022,Japanese-arpes,WrayARPES_PhysRevLett.124.076401,Thomas2020,li2021magnetic,Aoki2021,Ishizuka2021,Shishidou2021PRB,ShickPhysRevB.103.125136,LiuPRB.106.L241111,wilhelm2023xray,HazraPRL2023,KhmelevskyiPRB2023,ute2valence_MPI24-PRR.6.033299}. Furthermore, this suggests that under increasing pressure the Kondo mechanism explains both the observed decrease in \(H_m\) and, within our model, the phase transition into SC2 with increasing \(T_c^{\mathrm{(SC2)}}\).

\section*{Evolution of RKKY interactions}

The temperature dependence of the MM signal in the torque magnetometry, as well as magnetoconductivity data, reveals the complexity of the transition, of such subtlety that it proved challenging to resolve in pulsed field measurements \cite{Miyake2019,Miyake2021,qcl}. In particular, the intriguing dependence of some features on temperature and independence of others points to the presence of two coupled degrees of freedom, each possessing distinct behavior. 

Heavy-fermion systems are renowned for their rich phenomenology that stems from the coupling of local and itinerant degrees of freedom \cite{Coleman2015book}. Perhaps the most straightforward scenario is when the local moments fully participate in the Fermi surface and the system behaves as a Fermi liquid, albeit with renormalized parameters. The more intriguing possibility is when a residual local moment remains, be it because only some moments participate in the Fermi surface, or a gain in magnetic energy initiates a partial relocalization of the $f$-electrons at lower temperatures. This two-fluid composite of local moments and a heavy-fermion liquid possesses unique properties because some of the originally local degrees of freedom become non-local, whereas others remain so, and there are strong interactions that tie them together. A variety of phenomena can be exhibited as a result, such as easy-axis reorientation \cite{Scott2026}, or metamagnetic transitions \cite{mccollam2021lifshitz}. This possibility is particularly likely in multi-$f$-electron systems containing e.g. uranium, in which some $f$-electrons participate in the Fermi surface, whereas others retain their local-moment character. The strong interactions that originally act to tie them together, such as Hund's coupling or spin-orbit induced anisotropy, now give rise to strong and anisotropic interactions between the residual moment and the heavy-fermion fluid.

Given the macroscopic degeneracy of any local moments that remain, it is important to consider interactions within this subspace. The work of Ref. \cite{Scott2026} showed how an effective Hamiltonian can be derived for the local-moment subspace by integrating out the heavy-fermions. This Hamiltonian is essentially the RKKY exchange mediated by the heavy-fermions but renormalized by strong local interactions between the $f$-electrons that remain as local moments and those that participate in the Fermi surface by forming part of the heavy-fermion quasiparticle.

Considering, for simplicity, two $f$-electrons on each uranium ion and a general orthorhombic anisotropy for the total moment, the RKKY interaction that we obtain between the $f$-moments that remain if only one of the electrons participates in the Fermi surface is given by
\begin{align}
H_{\rm eff}\! =\!-\frac{1}{2}\sum_{i\neq  j ,\eta}&\left(  J_K^2\chi_{cc} (\mathbf{r}_i-\mathbf{r}_j) +J_KD^{\eta}\chi_{cf} (\mathbf{r}_i-\mathbf{r}_j)\right) 
\\\nonumber
&\times S^{\eta}_{}(\mathbf{r}_i) S^{\eta}_{}(\mathbf{r}_j),\! 
\end{align}
where $D^{\eta}=\{D^x,D^y,D^z\}$ are general anisotropy parameters, $J_K$ is the strength of the Kondo coupling and $S^{\eta}_{}(\mathbf{r}_i)$ is the $\eta$-component of the spin of the $f$-moment. $\chi_{cc}$ is the susceptibility of the conduction electrons when a field is only applied to them, whereas $\chi_{cf}$ is the susceptibility of the conduction electrons when the field is only applied to the hybridizing $f$-electrons.
We thus notice that the isotropic part of the RKKY interaction is controlled by $\chi_{cc}$ whereas the anisotropic part is controlled by $\chi_{cf}$. It has been recognized through extensive Knight shift studies \cite{TokunagaJPSJ.88.073701,AzariSonierPRB23,AzariPRB25} that $\chi_{cc}$ has weak dependence on temperature, whereas the $\chi_{cf}$ component -- which also measures the size of the coherent Fermi liquid -- has a strong temperature dependence below the coherence temperature $T^*_{\rm coh}$: 
\begin{align}
    \chi_{cf}(T)\propto\left(1-\frac{T}{T^*_{\rm coh}} \right) \log\frac{T}{T^*_{\rm coh}},
\end{align}
where we have taken the uniform part $\chi_{cf}=\sum_{ \{i | \mathbf{r}\neq \mathbf{0}\}}\chi_{cf}(\mathbf{r})$.

Note that for a given crystal electric field (CEF) anisotropy (i.e. fixed $D^{\gamma}$), the anisotropy of the exchange could oppose that of CEF or align with it depending on the sign of $\chi_{cf}$. Mean-field calculations have shown that beyond its steep rise below the coherence temperature $\chi_{cf}$ also changes sign around $T^*_{\rm coh}$, which could be responsible for the reorientation of the easy-axis that is seen across many heavy-fermion compounds \cite{Scott2026}. Indeed, such a reorientation from the $b$-axis at higher temperatures to the $a$-axis at lower temperatures is seen in UTe$_2$ at a temperature that tracks $T^*_{\rm coh}$~\cite{Aoki_UTe2review2022}.

Furthermore, $\chi_{cf}$ is highly sensitive to the level of hybridization and again changes sign as we tune from more localized magnets to more itinerant ones (stronger hybridization and dominance of the coherence energy over the RKKY energy $T^*_{\rm coh}\gg T_{\rm RKKY}$). In particular, Ref. \cite{Scott2026} found that magnetic order along the CEF hard-axis takes place in more itinerant systems, whereas magnetic order along the CEF easy-axis takes place in more localized magnets (lower ratio of coherence to RKKY energy). We can thus see that the anisotropy of the RKKY interaction is very sensitive to changes in temperature as well as the degree of localization. We stress here that the evolution of $\chi_{cf}(T/T^*_{\rm coh})$ is smooth and does not involve any Fermi surface reconstruction.

Increasing pressure is believed to tune UTe$_2$ from a more itinerant magnet to a more localized one, as supported by the experimental observation of a higher $f$ contribution to the Fermi surface with increasing pressure~\cite{weinberger2025pressure}, with Kondo screening likely suppressed with pressure. It is important to realize that this is a spectrum and UTe$_2$ should not be viewed as completely one or the other. With pressure, it is believed that it moves closer to $5f^2$ valency (further away from $5f^3$), which is generally thought to result in weaker hybridization with the conduction band and lower effective Kondo coupling, i.e., stronger localization \cite{Amorese2020, Deng2024}. 
Tuning pressure reduces $T^*_{\rm coh}$, Kondo hybridization and the level of itinerancy, and the system is nearing the point where $T_{\rm RKKY} \gtrsim T^*_{\rm coh}$ and magnetic order is favored over Kondo hybridization as per the Doniach phase diagram (barring significant Fermi surface reconstruction). The fact that the wavevector of AFM order seen above $p_c$ is not too far from the resonance seen in inelastic neutron scattering at ambient pressure~\cite{Duan2021AFMfluc} suggests that the spatial dependence of RKKY does not change and there is no significant Fermi surface reconstruction, which has been verified in moderate field strengths~\cite{weinberger2025pressure}. This is in keeping with $\chi_{cc}$, which sets the spatial dependence of the RKKY, being less sensitive to changes in Kondo coupling. However, the spin-space anisotropy of RKKY interactions, which is a lot more sensitive to changes in Kondo coupling, is likely to evolve strongly. In particular, $\chi_{cf}(T=0)$ is expected to decay quickly as pressure is lowered due to a falling Kondo coupling and $T^*_{\rm coh}$, which would result in more isotropic RKKY interactions on approaching $p_c$. With a reduced anisotropy between $a$- and $b$-directions, the cost of longitudinal fluctuations along $b$, proposed to drive SC2 \cite{tony2024enhanced}, would fall, thereby stabilizing the phase. $\chi_{cf}$ increases with cooling, leading to more costly fluctuations along the hard $b$-axis, which might explain the emergence of SC1 at the lowest temperatures. 

Recent large-$N$ studies have shown that $\chi_{cf}(T/T^*_{\rm coh},h/h^*)$ not only has a strong temperature dependence but also has an abrupt evolution with applied field at the lowest temperatures, experiencing a jump and change of sign at $h=h^*$, which results in a sudden reversal of the RKKY anisotropy. The jump in $\chi_{cf}$ may be caused by a Fermi surface reconstruction at large magnetizations~\cite{lewin2023review}, whereby the bands become more insulating and less compressible. This is a likely mechanism for the metamagnetic transition taking place at large $h_b$, with positive feedback leading to a runaway growth of longitudinal susceptibility. As the field grows and the magnetization grows with it, the RKKY exchange aniostropy begins to change abruptly, increasing the longitudinal susceptibility and making the magnetization grow even faster. Negative feedback takes place when the field is applied to the easy-axis. In this case, as the magnetization grows the longitudinal susceptibility falls and reduces the fall of magnetization, resulting in its rapid saturation.

The above mechanism relies on feedback between the conduction band and the polarization of the local moments, which could be behind the intriguing temperature dependence of torque magnetometry as well as magnetoconductivity data. While the local moments polarize suddenly as the critical RKKY anisotropy and field are reached -- leading to the sharpest and temperature-independent features of the field profile of the MM transition -- the response of the band shifts and broadens, at only slightly elevated temperatures, leading to the gentler and more temperature-dependent features of the field profile. This is because  $\chi_{cf}(T/T^*,h/h^*)$ becomes broader as a function of $h$ when temperature is increased, while the critical field $h^*$ remains approximately constant.


\section*{Discussion}

The fine structure resolved within the UTe$_2$ $\tau(H)$ profile at the metamagnetic phase boundary (Fig.~\ref{fig:magnetisation}b) demonstrates that this field-induced transition is more complex than a simple, singular process. Furthermore, the positive gradient of $M(H)$ after the transition (Fig.~\ref{fig:magnetisation}a), up to at least 80~T, clearly shows that there is still a considerable susceptibility of the itinerant moments in the PPM state.

Metamagnetic Sr$_3$Ru$_2$O$_7$~\cite{lester2015field,lester2021magnetic} and URu$_2$Si$_2$~\cite{knafo2016field} both possess spin-density wave (SDW) phases nestled against their field-polarized first order metamagnetic phase boundaries. In light of our $\tau(H)$ data, similar phenomenology in UTe$_2$ appears to be highly plausible. Angle-dependent $\tau(H)$ measurements, to resolve the anisotropic evolution of separate components of \textbf{M} contributing to the overall signal (through $\mathbf{M} \cross \mathbf{B}$), are required to understand this better. We posit that it appears likely that such a SDW phase might be the origin for the strange metallicity observed where SC3 exhibits its maximum critical temperature and upper critical field values~\cite{weinberger2025strangemetallicityencompasseshigh}. This magnetic field tilt angle (of around 34$\degree$ from \textit{b} towards \textit{c}) is also the section of the $b-c$ plane in which the onset of SC3 has been observed to extend down to the lowest $H$ value below the metamagnetic transition~\cite{wu2026directobservationspilloverhigh}. A quantum critical point marking the termination  of SDW order thus appears a likely candidate for the pairing glue underpinning SC3.

We note that no signatures of Shubnikov-de Haas quantum oscillations were resolved in our contactless resistivity measurements at high fields; nor were any signatures of non-Onsager magneto-oscillations such as from the Shiba-Fukuyama-Stark quantum interference effect~\cite{shiba1969,stark1971quantum}, which have previously been resolved in UTe$_2$~\cite{BroylesPRL23,theo2024,HusstedtPRB25,weinberger2025pressure} (albeit under different field orientations than for \textbf{H} $\parallel b$ studied here). This is despite the low measurement temperatures, large applied magnetic field strengths, and high crystalline quality of the studied TDO sample (from the same growth batch as prior quantum oscillatory studies~\cite{Eaton2024,theo2024}).

In the superconducting states, the Fermi surface is gapped out. Therefore, we should only expect to resolve quantum oscillations, if present, within the non-superconducting phases accessed in our TDO study (i.e. the PPM and AFM states, at high fields/pressures). We posit there are several possible reasons for why no oscillations were resolved: firstly, for the AFM state, this phase terminates at $\approx$~10~T, which is a relatively modest upper boundary for attempting to resolve any oscillatory features that may be present (at lower fields). Furthermore, for the PPM phase, it is possible that (i) the mean free path of the studied sample was simply insufficient; and/or (ii) that the Fermi surface in the PPM state, similar to the normal paramagnetic state~\cite{AokidHvA_UTe2-2022,Eaton2024}, is quasi-2D such that no closed orbits are expected for \textbf{H} $\parallel b$; and/or (iii) that the effective masses become so large at high pressure that the measurement temperature was too high to resolve coherent Landau quantization. Support for scenario (iii) comes from several resistivity~\cite{Braithwaite2019,Thomas2020,Valiska2021,KnebelPRBcaxis24,kim2025tuning} and specific heat~\cite{vasina2026quantitativethermodynamicstudysuperconducting} studies under pressure, which discern a marked increase in the quasiparticle effective mass by a factor of $\approx$ 3 in the vicinity of $p_c$, compared to at $p = 0$.

In a previous ambient pressure study of SC2~\cite{tony2024enhanced}, we proposed a model for equal-spin pair formation underpinning this state due to metamagnetic fluctuations emanating from $H_m$. Experimental evidence corroborating this scenario has been observed in measurements of NMR~\cite{Kinjo23,tokunaga2023longitudinal} and of the magnetic entropy~\cite{TokiwaPRB2024}, while ultrasound measurements under pressure have confirmed the presence of strong fluctuations~\cite{kamat2026vanishingphasestiffnessfluctuationdominated}. In the present work, we extended this model to include the response under hydrostatic pressure. Upon increasing $p$ towards $p_c$, $H_m$ is observed to decrease monotonically (Fig.~\ref{fig:pressures})~\cite{LinNevidomskyyPaglione20}. Meanwhile, SC2 is stabilized in ambient $H$ for $p\gtrapprox$~3~kbar, with its critical temperature then steadily rising until dropping off close to $p_c$~\cite{tony25pressure,Thomas2020,Braithwaite2019}. At 70~mK, we find that the transition field between SC1-SC2, for \textbf{H} $\parallel b$, monotonically decreases as $p$ is increased towards $p_c$ (Fig.~\ref{fig:pressures}b).

We modeled this behavior theoretically -- of shrinking $H_m$ and strengthening SC2 with increasing $p$ -- by considering the evolution of Kondo and RKKY interactions as $p \to p_c$. We propose that upon approaching $p_c$, a decaying Kondo coupling reduces the spin-space anisotropy between the $a$ and $b$ crystallographic directions. This reduction in magnetic anisotropy significantly lowers the energy cost of longitudinal spin fluctuations along the hard $b$ axis. If our model~\cite{tony2024enhanced} is correct in its attribution of the formation of the SC2 state to the presence of these $Q=0$ fluctuations, then their anticipated proliferation under pressure provides a natural explanation for why increasing compression stabilizes the SC2 state and enhances its critical temperature.

\vspace{-0mm}
\begin{acknowledgments}\vspace{-5mm}
We are grateful to A.J. Hickey and G.G. Lonzarich for stimulating discussions. This project was supported by the EPSRC of the UK (grants EP/Z533695/1, EP/X011992/1 \& EP/R513180/1). Crystal growth and characterization were performed in MGML (mgml.eu), which is supported within the program of Czech Research Infrastructures (project no. LM2023065). We acknowledge financial support by the Czech Science Foundation GAČR under the Junior Star Grant No. 26-21795M (STiUS). A portion of this work was performed at the National High Magnetic Field Laboratory, which is supported by National Science Foundation Cooperative Agreement No. DMR-2128556 and the State of Florida. A portion of this work was carried out at the Synergetic Extreme Condition User Facility (SECUF \href{https://cstr.cn/31123.02.SECUF}{https://cstr.cn/31123.02.SECUF}). This work was performed in part at the Aspen Center for Physics, which is supported by National Science Foundation grant PHY-2210452. T.I.W. acknowledges support from Murray Edwards College (University of Cambridge) and the Cambridge Philosophical Society through a Henslow Fellowship. T.I.W. and A.G.E. acknowledge support from QuantEmX grants from ICAM and the Gordon and Betty Moore Foundation through Grants GBMF5305 \& GBMF9616 and from the US National Science Foundation Grant Number 2201516 under the Accelnet program of Office of International Science and Engineering. The work of D.S. was financially supported by the NSF Quantum Leap Challenge Institute for Hybrid Quantum Architectures and Networks Grant No. OMA-2016136. D.V.C. acknowledges financial support from the National High Magnetic Field Laboratory through a Dirac Fellowship, which is funded by the National Science Foundation (Grant No. DMR-1644779) and the State of Florida, and from Washington University in St. Louis through the Edwin Thompson Jaynes Postdoctoral Fellowship. A.G.E. acknowledges support from the Henry Royce Institute for Advanced Materials through the Equipment Access Scheme enabling access to the Advanced Materials Characterisation Suite at Cambridge, grants EP/P024947/1, EP/M000524/1 \& EP/R00661X/1; and from Sidney Sussex College (University of Cambridge).
\end{acknowledgments}

\bibliography{UTe2}
\end{document}